*Journal of Physics: Condensed Matter*: Accepted, November 25, 2025

# Revising the Theory of Cross Phenomena

Zi-Kui Liu*

Department of Materials Science and Engineering, The Pennsylvania State University,

University Park, Pennsylvania 16802, USA

* Corresponding Author Email: prof.zikui.liu@psu.edu

**Abstract**

In the present paper, the theory of cross phenomena (TCP) proposed by the present author is revised by starting from the first law of thermodynamics instead of the combined law of thermodynamics in the previous publication because the potentials in the latter are not independent of each other. TCP aims to derive the flux equations with respect to independent driving forces from the fundamental laws of thermodynamics rather than the phenomenological Onsager flux equations, which are flawed as shown by the revised TCP.

## 1 Introduction

Recently, the present author proposed a theory of cross phenomena (TCP) [1–4], which asserts that the flux of a molar quantity is governed solely by the gradient of its conjugate potential. This contrasts with the classical Onsager theorem, i.e., flux equations and reciprocal relations [5,6], which posit that fluxes are linearly related to the gradients of all potentials through a symmetric kinetic coefficient matrix. The central question was that if the symmetric matrix is diagonalized, what are the driving forces, and can these driving forces be directly derived from fundamental laws of thermodynamics? Unfortunately, the present author started from the combined law of thermodynamics and failed to realize that the potentials in the combined laws are not independent of each other even though the dependences were derived previously by the present author [7,8] and revisited recently [9].

In the present work, the original TCP is revised with the flux equation derived from the first law of thermodynamics and the definitions of the total entropy change, the total work change, and the chemical potential. Although the revision is foundational, both original and revised TCP lead to similar conclusions with respect to the Onsager theorem. Since the present work focuses on revising the previous TCP by the present author, the references on Onsager theorem in the literature are not extensively listed, and readers are referred to the citations in previous publications [1–4].

It is important to note that the Onsager theorem is phenomenological in nature. Hillert [10] emphasized this point, and Balluffi et al. [11] remarked that Onsager's relations are not as fundamental as the first and second laws of thermodynamics. This view is consistent with Onsager's own statement [5] that "the principle of microscopic reversibility is less general than the fundamental laws of thermodynamics." Other related discussions can be found in the literature [12–19].

The original formulation of TCP aimed to derive flux equations directly from the fundamental laws of thermodynamics. However, it did not adequately account for the interdependence of molar quantities inherent in the combined law of thermodynamics, resulting in explicit relation between potentials [7–9] as shown by Eq. 2 below in this paper, and thus incorrect flux equations in previous

TCP [1–4]. To correct this mistake, the present paper revises TCP by deriving the flux equations starting from the first law of thermodynamics, rather than the combined law. This approach ensures a consistent and fundamental thermodynamic foundation for TCP.

Recently, Chen and Mauro [20] published a study on entropy and mass transport in thermoelectric and thermochemical systems. They demonstrated that entropy flux is driven by gradients in both temperature and the chemical potential of electrons or components, respectively, through mechanistical considerations. Furthermore, they showed that the diffusion flux of a component is proportional to the gradients of both its conjugate chemical potential and temperature, but involving only a single kinetic coefficient for each chemical species, rather than two if Onsager theorem would be applied. Furthermore, they did not address diffusion in multicomponent systems in terms of Onsager theorem. Inspired by their insights, the present work revises the previously proposed TCP to correct the conceptual and mathematical errors in the earlier formulation [1–4].

## 2 Brief review of previous TCP and its inconsistence

The present author previously introduced the TCP within the framework of Ågren's atomic mobility formalism, formulated in the *lattice-fixed frame of reference* [21–23] as shown in Figure 1. In that context, four critical questions were raised concerning the Onsager theorem [1–4]. The central question was: Given that a symmetric matrix can be diagonalized, what is the form of the driving force vector after diagonalizing the symmetric kinetic coefficient matrix in the Onsager formalism? Furthermore, can this driving force vector be derived directly from the fundamental laws of thermodynamics, thereby avoiding the limitations of phenomenological assumptions and potential misinterpretation based on experimental observations?

Under isothermal and isobaric conditions, Ågren and his collaborators [19,21–23] demonstrated that, in multicomponent diffusion systems, the flux of component $i$ in the lattice-fixed frame of reference, $J_i$, is driven by the gradient of its chemical potential, $\nabla\mu_i$. In this formulation, the kinetic coefficient matrix becomes diagonal, denoted by $L_i$ in Figure 1, the flux equation can be expressed accordingly [19]

$$J_i = -L_i\nabla\mu_i = -L_i\sum\frac{\partial\mu_i}{\partial c_j}\nabla c_j = -\sum {}^{I}D_{ij}\nabla c_j \qquad \textit{Eq. 1}$$

where $c_j$ is the concentration of component $j$, and ${}^{I}D_{ij} = L_i\frac{\partial\mu_i}{\partial c_j}$ the intrinsic diffusivity of component $i$ with respect to the concentration gradient of component $j$ as shown in Figure 1. The negative sign reflects the direction of flux being down the potential gradient, consistent with dissipative processes. The off-diagonal intrinsic diffusion coefficients are due to the dependence of $\mu_i$ on the concentrations of all components, i.e., the thermodynamic factor as the second partial derivative of Gibbs energy with respect to $c_i$ and $c_j$, $\frac{\partial^2 G}{\partial c_i\partial c_j}$. Notably, there is only one kinetic coefficient, $L_i$, for each component, emphasizing the diagonal nature of the kinetic matrix in this formalism, and the intrinsic diffusivity matrix is not symmetrical due to $L_i \neq L_j$, $\frac{\partial\mu_i}{\partial c_j} \neq \frac{\partial\mu_j}{\partial c_i}$, and ${}^{I}D_{ij} \neq {}^{I}D_{ji}$ in general.

Additional types of off-diagonal diffusion coefficients can be introduced by transforming the flux description to the *volume-fixed frame of reference*, i.e. as represented by $L'_{ij}$ in Figure 1, while keeping chemical potential gradients as the driving force. In this frame, the chemical potential gradient is expanded into the gradients of all component concentrations in the system, as expressed by the summation in Eq. 1, This leads to the formulation of interdiffusion diffusivity matrices resulting in the interdiffusion diffusivity matrixes, $D_{ij}$, under *volume-fixed frames of reference*, also illustrated in Figure 1, which is not symmetrical in general.

It is thus evident that the driving forces and the frame of reference must be defined simultaneously to ensure the proper definition of fluxes and then the entropy productions due to the internal processes. While in the literature the clear definition of the frame of reference was often not explicitly specified [20,24,25]. It should be mentioned that ${}^{I}D_{ij}^{n}$ and $D_{ij}^{n}$ in Figure 1 denote the reduced intrinsic and interdiffusion coefficients due to the mole fractions of all components summed to unit where $n$ represents the reference component. It is thus natural that some of the diffusivity coefficients in the ${}^{I}D_{ij}^{n}$ and $D_{ij}^{n}$ matrixes may be negative including those with $i = j$ [26], depending on which component is selected as the reference component.

Experimental measurements typically correspond to the interdiffusion diffusivity matrices, except for tracer diffusivities ($D_i^*$), determined using isotopic labeling, and intrinsic diffusivities, obtained using internal markers [27–29]. To inversely determine the kinetic coefficient or the atomic mobility as functions of temperature and composition, the Gibbs energy is required. This approach forms the basis for developing atomic mobility databases in multicomponent systems [30]. It is also noteworthy that the widely used Maxwell–Stefan diffusion equations, formulated in terms of concentration gradients, involve (c−1)×(c−1) diffusion coefficients for a system with c components [31,32], which are $D_{ij}^n$ in Ågren's atomic mobility formalism as shown in Figure 1 in the volume-fixed reference.

Furthermore, Ågren's formalism elegantly accounts for uphill diffusion, wherein a component migrates from regions of its lower concentration (corresponding to higher chemical potential) to regions of its higher concentration (lower chemical potential). A classic example is the uphill diffusion of carbon in steels, driven by the inhomogeneous distribution of silicon. As observed by Darken [33], high silicon concentrations significantly increase the chemical potential of carbon, resulting in a flux of carbon ($J_C$) that is in the same direction as its concentration gradient ($\nabla c_C$) as described in Eq. 1. The Kirkendall effect, manifested through marker migration and void formation [27–29], has also been successfully predicted using numerical simulation tools that incorporate Gibbs energy functions and atomic mobility databases [34]. These simulations provide a powerful framework for understanding diffusion phenomena in multicomponent systems where the combination of kinetic coefficients and thermodynamic factors results in rich diffusion behaviors.

The present author subsequently extended Ågren's approach to incorporate temperature gradients in the development of TCP [1–4]. However, this previous formulation was based on the combined law of thermodynamics and failed including the *interdependence* of entropy and mass changes as previously defined by the present author [7,8]. This introduced a conceptual inconsistency in the derivation of flux equations for *independent* internal processes. Thanks to the insights provided by Chen and Mauro [20], this issue is now addressed, and a revised formulation of TCP is presented in the current work.

## 3 Revision of TCP

As will be demonstrated below, it is essential to begin with the first and second laws of thermodynamics to ensure that the driving forces for fluxes are treated in terms of *independent* variables. Accordingly, the first, second, and combined laws of thermodynamics [8,9] are represented in Table 1 in the context of this revised theory with all the symbols defined in the footnote of the table.

Table 1 highlights two important features previously discussed in Ref. [7–9], which are worth emphasizing here. The first concerns the definition of potentials. As articulated by Narasimhan [35], "potential theory involves problems describable in terms of a partial differential equation, in which the dependent variable is the appropriate potential." In Table 1, $U_i$, $T$, and $-P$ are defined as the partial derivatives of internal energy with respect to $N_i$, $S$, and $V$, respectively, are thus potentials, and molar energy and molar free energies are not potentials.

The second feature is the distinction between $U_i$ and $\mu_i$. Although both are partial derivatives of internal energy with respect to $N_i$, $\mu_i$ is taken under constant entropy and volume (i.e., S and V held fixed), while $U_i$ under adiabatic and no work exchange conditions. As such, $\mu_i$ is related to the other three potentials through the relation shown in the final equation in the footnote of Table 1, which is repeated below due to its fundamental importance,

$$\mu_i = U_i - TS_i + PV_i = \left(\frac{\partial U}{\partial N_i}\right)_{S,V,N_{j\neq i},d_{ip}S=0} \qquad Eq.\ 2$$

Eq. 2 originates from the contributions of changes in $N_i$ to the entropy and volume of the system, as discussed in Refs. [7–9]. Its significance will become more evident in the subsequent discussion on mass and entropy transport presented in the present paper. Since $U_i$ is defined under conditions where there is no exchange of heat or work, it captures the intrinsic contribution of mass exchange to the internal energy. In this sense, $U_i$ is more general than the chemical potential $\mu_i$, which is defined under constant entropy and volume and thus inherently includes the effects of heat and work exchanges between the system and its surroundings, both of which depend on mass exchange.

According to the first law of thermodynamics, the exchanges of heat, work, and mass between a system and its surroundings, denoted as $dQ$, $dW$, and $dN_i$, respectively, can be independently controlled from the surroundings. In contrast, the combined law of thermodynamics introduces a coupling between entropy and mass changes, such that $dS$ and $dN_i$ are no longer independent, as expressed by the entropy change equation shown in Table 1. As a result, the fluxes of entropy and components are inherently interdependent. This interdependence persists even when various forms of work are considered, due to the contribution of each component's partial molar volume [9]. Therefore, it is advantageous to base the analysis of heat and mass fluxes on the first law of thermodynamics, where the transport of each component can be treated as independent. In the present work, all fluxes are defined in *the lattice-fixed frame of reference*, following the formalism established by Agren in Refs. [19,21–23].

The heat flux $J_Q$ can be expressed using Fourier's law [35], as shown below and illustrated schematically in Figure 2:

$$J_Q = -L_Q \frac{T_L - T_H}{\Delta z} = -L_Q \nabla T \qquad \textit{Eq. 3}$$

where $L_Q$ is the kinetic coefficient for heat conduction (i.e., the thermal conductivity), $T_H$ and $T_L$ are the high (left in Figure 2) and low (right in Figure 2) temperatures, respectively, and $\Delta z$ is the unit distance over which heat is transported.

The entropy production $d_{ip}S$ due to heat conduction can be calculated using the definition of entropy production provided in Table 1:

$$d_{ip}S = \frac{dQ}{T_L} - \frac{dQ}{T_H} = \frac{dQ}{T_L T_H}(T_H - T_L) \geq 0 \qquad \textit{Eq. 4}$$

This expression confirms compliance with the second law of thermodynamics, as the entropy production is always non-negative.

The rate of internal energy change due to the transport of component $i$ in the system is schematically illustrated in Figure 3 and can be expressed as:

$$\frac{dU}{A\Delta z dt} = \dot{U}_V = \frac{dN_i}{Adt}\frac{U_i^L - U_i^H}{\Delta z} = \frac{dN_i}{Adt}\frac{\Delta U_i}{\Delta z} = J_i \nabla U_i \qquad \textit{Eq. 5}$$

$$J_i = \frac{dN_i}{Adt} \qquad \textit{Eq. 6}$$

where $A$ and $dt$ are the cross-sectional area and time interval of mass transport, $\dot{U}_V$ is the volumetric rate of internal energy change, $J_i$ is the flux of component $i$, $U_i^L$ and $U_i^H$ are the partial internal energies of component $i$ on the low (right) and high (left) sides of the transport direction, and $\Delta U_i$ and $\nabla U_i$ are the difference and the gradient of the partial internal energies of component $i$ along the transport direction, respectively.

The central question now is how to define the flux $J_i$ of component $i$. Since the change of each component is *independent*, and only the product of $dN_i$ and $U_i$ contributes to the internal energy change in the first law of thermodynamics (as shown in Table 1), the flux of component $i$ is postulated to be proportional *solely* to its potential gradient, $\nabla U_i$, as follows

$$J_i = \frac{dN_i}{Adt} = -L_i \nabla U_i \qquad \textit{Eq. 7}$$

where $\nabla U_i$ includes contributions from the spatial variations of all independent variables involved in the internal process. This formulation parallels Fourier's law for heat conduction (as shown by Eq. 3), but is grounded directly in the first law of thermodynamics, offering a more fundamental perspective. Substituting Eq. 7 into Eq. 5 yields

$$\dot{U}_V = J_i \nabla U_i = -L_i (\nabla U_i)^2 \leq 0 \qquad \textit{Eq. 8}$$

which demonstrates that mass transport results in a local decrease in internal energy density, reflecting the system's intrinsic drive to redistribute internal energy as it progresses toward equilibrium through internal processes.

On the other hand, according to Onsager's theorem, the flux $J_i$ also depends on the gradients of partial internal energies of other components, i.e., $\nabla U_j$, with an independent kinetic coefficient, which would lead to cross terms such as $\nabla U_i \nabla U_j$ in Eq. 8. This introduces coupling between independent components and undermines the independence of mass exchange processes as defined by the first law of thermodynamics. The independent kinetic coefficient implies that the component $i$ has different diffusion mechanisms with respect to every other component, which is physically unrealistic. Furthermore, since both $\nabla U_i$ and $\nabla U_j$ incorporate contributions from spatial variations of all independent variables associated with the internal process, the cross term $\nabla U_i \nabla U_j$ may lead to an overrepresentation of the influence of independent variables on $J_i$. As discussed earlier in the context of atomic mobility and various diffusion coefficients, such cross terms can arise either from a change in the frame of reference or from explicitly expressing $\nabla U_i$ in terms of the gradients of all independent variables, both of which reveal the underlying coupling mechanisms. They form the foundation of the TCP framework introduced in Refs. [1–4] and further revised in the remainder of the present paper.

According to the first law of thermodynamics, heat and mass transport are two independent processes, represented by Eq. 3 and Eq. 7, respectively. As a result, the kinetic coefficient matrix is diagonal, as shown in the matrix form of the flux expression below:

$$\begin{pmatrix} J_Q \\ J_1 \\ J_2 \\ \vdots \\ J_i \\ \vdots \\ J_c \end{pmatrix} = -\begin{pmatrix} L_Q & 0 & 0 & \cdots & 0 & \cdots & 0 \\ 0 & L_1 & 0 & \cdots & 0 & \cdots & 0 \\ 0 & 0 & L_2 & \cdots & 0 & \cdots & 0 \\ & & & \cdots & & & \\ 0 & 0 & 0 & \cdots & L_i & \cdots & 0 \\ & & & \cdots & & & \\ 0 & 0 & 0 & \cdots & 0 & \cdots & L_c \end{pmatrix} \begin{pmatrix} \nabla T \\ \nabla U_1 \\ \nabla U_2 \\ \vdots \\ \nabla U_i \\ \vdots \\ \nabla U_c \end{pmatrix} \qquad \textit{Eq. 9}$$

where $c$ is the number of independent components in the system, and the total number of independent kinetic coefficients is $c + 1$.

However, this independence no longer holds when entropy flow is considered. In the presence of both heat conduction and mass transport, the total entropy change, assuming that the configurational entropy production is stationary ($d_{ip}S^{conf}/dt = 0$), the entropy flow $J_S$ can be expressed, based on Table 1, as

$$dS = \frac{dQ}{T} + \sum S_i dN_i \qquad \textit{Eq. 10}$$

$$J_S = \frac{dS}{Adt} = \frac{J_Q}{T} + \sum S_i J_i = -\frac{L_Q}{T}\nabla T - \sum S_i L_i \nabla U_i \qquad \textit{Eq. 11}$$

where subscript *ip* in Eq. 10 and Eq. 11 is omitted for simplicity. The middle section of Eq. 11 is obtained by the differentiation of Eq. 10 with respect to time. These expressions show that entropy flow is intrinsically linked to both heat conduction and the mass transport of all components, highlighting their coupled nature.

Let us now consider the combined law of thermodynamics, incorporating both entropy and chemical potential. The relationship among the potentials is given by Eq. 2, and the gradient of partial internal energy and fluxes of mass and entropy can be expressed, respectively, as follows:

$$\nabla U_i = \nabla \mu_i + S_i \nabla T - V_i \nabla P \qquad \textit{Eq. 12}$$

$$J_i = -L_i \nabla U_i = -L_i(\nabla \mu_i + S_i \nabla T - V_i \nabla P) \qquad \textit{Eq. 13}$$

$$J_S = -\frac{L_Q}{T}\nabla T - \sum S_i L_i(\nabla \mu_i + S_i \nabla T - V_i \nabla P) \qquad \textit{Eq. 14}$$

Under isobaric condition ($\nabla P$=0), Eq. 13 corresponds to Eq. 38 by Chen and Mauro [20]. Under *isothermal* and *isobaric* conditions ($\nabla T$=0 and $\nabla P$=0), it reduces to Eq. 1 as formulated by Agren and his collaborators [19,21–23]. While the dependence of $\nabla \mu_i$ on concentration is captured by Eq. 1, it should not be extended to $\nabla T$ or $\nabla P$ as in the previous TCP framework [1–4], since these contributions are already explicitly included in Eq. 12 via Eq. 2. Nevertheless, it is important to recognize that the concentration gradients, $\nabla c_j$ in Eq. 1, are themselves influenced by $\nabla T$ and $\nabla P$.

It is thus evident that the experimentally measured Seebeck and Soret coefficients, corresponding to the coefficients of $\nabla T$ in Eq. 13 for electrons and chemical components, respectively, represent the partial entropies of electrons and chemical species. These coefficients can be predicted from thermodynamic principles, as demonstrated in the publications by the present author's team and collaborators[36,37]. Experimentally, they are typically evaluated under steady-state conditions, i.e., $J_i = 0$, leading to the following expression for the Seebeck or Soret coefficient, collectively denoted as $S_T$,

$$S_T = -\left(\frac{\partial \mu_i}{\partial T}\right) \qquad \textit{Eq. 15}$$

This expression highlights the thermodynamic origin of $S_T$ as a temperature derivative of the chemical potential. It is thus evident that they are not kinetic coefficients, and the only independent kinetic coefficient for component $i$ is $L_i$ shown in Eq. 13. It is noted that in the literature, the Seebeck and Soret coefficients are usually scaled by the elementary charge and by the product of Boltzmann's constant and temperature, respectively. To evaluate $\mu_i$ and its derivatives under equilibrium and nonequilibrium conditions, the free energy landscape as a function of independent variables are needed, which can be predicted by the zentropy theory developed by the present author's team [3,38–40].

Accordingly, the matrix form of the flux equations for entropy and components with respect to $\nabla T$ and $\nabla \mu_i$, assuming $\nabla P = 0$ for simplicity, becomes

$$\begin{pmatrix} J_S \\ J_1 \\ J_2 \\ \vdots \\ J_i \\ \vdots \\ J_c \end{pmatrix} = -\begin{pmatrix} \frac{L_Q}{T} + \sum S_i{}^2 L_i & S_1 L_1 & S_2 L_2 & \cdots & S_i L_i & \cdots & S_c L_c \\ S_1 L_1 & L_1 & 0 & \cdots & 0 & \cdots & 0 \\ S_2 L_2 & 0 & L_2 & \cdots & 0 & \cdots & 0 \\ & & \cdots & & & & \\ S_i L_i & 0 & 0 & \cdots & L_i & \cdots & 0 \\ & & \cdots & & & & \\ S_c L_c & 0 & 0 & \cdots & 0 & \cdots & L_c \end{pmatrix} \begin{pmatrix} \nabla T \\ \nabla \mu_1 \\ \nabla \mu_2 \\ \vdots \\ \nabla \mu_i \\ \vdots \\ \nabla \mu_c \end{pmatrix} \qquad \textit{Eq. 16}$$

This formulation clearly shows that the coupling between entropy flow and mass transport arises from the partial entropies $S_i$, while the kinetic coefficients $L_i$ and $L_Q$ remain the only independent transport parameters.

The kinetic coefficient matrix in Eq. 16 is symmetric and diagonal except the first row and first column. Importantly, the number of independent kinetic coefficients remains unchanged, $c + 1$, as in the diagonal matrix of Eq. 9. This consistency arises because the first term in Eq. 13 originates from the first law of thermodynamics, as shown in Eq. 6, while the second term is derived from the definition of chemical potential provided in Table 1 and Eq. 2 and Eq. 12.

The revised TCP is therefore grounded in the first, second, and combined laws of thermodynamics. It is both fundamentally and mathematically distinct from the phenomenological Onsager theorem [10–19]. Notably, Eq. 2. and Eq. 12 were inadvertently omitted in the original development of TCP [1–4], which led to an incorrect representation of the coupling mechanisms.

Concentration gradients can be incorporated into the flux equations by combining Eq. 13 and Eq. 14 with Eq. 1, leading to the following expressions:

$$J_i = -L_i\left(\sum \frac{\partial \mu_i}{\partial c_j}\nabla c_j + S_i\nabla T - V_i\nabla P\right) \qquad \textit{Eq. 17}$$

$$J_S = -\frac{L_Q}{T}\nabla T - \sum S_i L_i\left(\sum \frac{\partial \mu_i}{\partial c_j}\nabla c_j + S_i\nabla T - V_i\nabla P\right) \qquad \textit{Eq. 18}$$

The corresponding kinetic coefficient matrix, incorporating the effects of concentration gradients, is then expressed as:

$$\begin{pmatrix} J_S \\ J_1 \\ J_2 \\ \vdots \\ J_i \\ \vdots \\ J_c \end{pmatrix} = -\begin{pmatrix} \frac{L_Q}{T} + \sum {S_i}^2 L_i & S_1L_1 & S_2L_2 \cdots & S_iL_i & \cdots S_cL_c \\ S_1L_1 & L_1\frac{\partial \mu_1}{\partial c_1} & L_1\frac{\partial \mu_1}{\partial c_2} \cdots & L_1\frac{\partial \mu_1}{\partial c_i} \cdots & L_1\frac{\partial \mu_1}{\partial c_c} \\ S_2L_2 & L_2\frac{\partial \mu_2}{\partial c_1} & L_2\frac{\partial \mu_2}{\partial c_2} \cdots & L_2\frac{\partial \mu_2}{\partial c_i} \cdots & L_2\frac{\partial \mu_2}{\partial c_c} \\ & & \cdots & & \\ & & \cdots & & \\ S_iL_i & L_i\frac{\partial \mu_i}{\partial c_1} & L_i\frac{\partial \mu_i}{\partial c_2} \cdots & L_i\frac{\partial \mu_i}{\partial c_i} \cdots & L_i\frac{\partial \mu_i}{\partial c_c} \\ & & \cdots & & \\ & & \cdots & & \\ S_cL_c & L_c\frac{\partial \mu_c}{\partial c_1} & L_c\frac{\partial \mu_c}{\partial c_2} \cdots & L_c\frac{\partial \mu_c}{\partial c_i} \cdots & L_c\frac{\partial \mu_c}{\partial c_c} \end{pmatrix}\begin{pmatrix} \nabla T \\ \nabla c_1 \\ \nabla c_2 \\ \vdots \\ \nabla c_i \\ \vdots \\ \nabla c_c \end{pmatrix} \qquad \textit{Eq. 19}$$

Since the kinetic coefficients $L_i$ are generally independent of one another, the resulting kinetic coefficient matrix is not symmetric in most cases. An important exception occurs in ionic solutions, where the migrations of cations and anions are often fully coupled, effectively behaving as a single transport process. This phenomenon has been discussed by Onsager [41] and the present author [42].

Furthermore, as recently discussed [9], for systems involving mechanical, electric, and magnetic work, the relationship among the relevant potentials can be expressed as:

$$U_i = \mu_i + S_iT - V(\sigma\varepsilon_i + E\theta_i + \mathcal{H}B_i) \qquad \textit{Eq. 20}$$

where $\sigma$, $E$, and $\mathcal{H}$ represent the stress, electric field, and magnetic field, and $\varepsilon_i$, $\theta_i$, and $B_i$ denote the partial strain, partial electric displacement, and partial magnetic flux associated with component $i$, respectively. These additional terms introduce further coupling mechanisms, and various cross phenomena can be derived by extending Eq. 13 as follows

$$J_i = -L_i \nabla U_i = -L_i \{\nabla \mu_i + S_i \nabla T - V(\varepsilon_i \nabla \sigma + \theta_i \nabla E + B_i \nabla \mathcal{H})\} \qquad \textit{Eq. 21}$$

A detailed treatment of these effects will be presented in future publications.

## 4 Summary

In summary, the revised TCP is derived directly from the first law of thermodynamics, followed by the definitions of entropy change, work exchange, and chemical potential for the combined law of thermodynamics. This differs from the previous TCP formulation, which was based on the combined law of thermodynamics and overlooked the intrinsic interdependence between entropy and mass changes, as well as between temperature and chemical potentials. Despite these foundational differences, both the original and revised TCP frameworks lead to similar conclusions, as shown below.

TCP demonstrates that the kinetic coefficient matrix governing heat and mass transport is diagonal in the *lattice-fixed frame of reference*, when expressed in terms of the gradients of temperature and the partial internal energies of individual components based on the first law of thermodynamics. By explicitly incorporating the dependence of entropy and volume changes on mass transport, TCP reveals that cross phenomena arise from the interdependence of potential gradients rather than from independently coupled fluxes, as postulated by the Onsager theorem.

As a result, the number of independent kinetic coefficients remains unchanged from that of the diagonal matrix. This stands in clear distinction to the Onsager framework, which introduces a larger number of seemingly independent coefficients through a phenomenological approach, further expanded in the Onsager–Stefan–Maxwell formalism. In contrast, TCP is firmly grounded in fundamental thermodynamic principles and avoids the phenomenological assumptions and the proliferation of coefficients that lack direct thermodynamic grounding in the Onsager theorem.

It is noted that the atomic mobility databases based on Agren's formalism as a function of composition and temperature have been developed for a large number of multi-component materials systems and broadly used worldwide, which can be used to directly calculate Soret coefficients through the partial entropy shown by Eq. 17. For Seebeck coefficients, the partial entropy of electrons are needed as shown by publications from the present author's group [36,37]. The databases for $L_Q$ can be developed similarly plus those for volumetric flow or plastic deformation [43], which will be discussed in the future.

**Acknowledgements**

The author is grateful to Long-Qing Chen for pointing out errors in early version of TCP and thanks John Mauro and David McDowell for valuable comments on the manuscript. This work was supported by the U.S. Department of Energy (DOE) through Grant No. DE-SC0023185.

*Table 1: First, second, and combined laws of thermodynamics*

| | | |
|---|---|---|
| First law of thermodynamics | $dU = dQ + dW + \sum U_i dN_i$ | ref. [8] |
| Second law of thermodynamics | $d_{ip}S = \frac{d_{ip}Q}{T} - \sum S_i^n dN_i^n + \sum S_j^w dN_j^w + d_{ip}S^{conf} \geq 0$ | |
| Entropy change of the system | $dS = \frac{dQ}{T} + \sum S_i dN_i + d_{ip}S$ | |
| Combined law of thermodynamics | $dU = TdS + dW + \sum (U_i - TS_i) dN_i - Td_{ip}S$ | |
| Combined law of thermodynamics with hydrostatic work | $dW = -P\left(dV - \sum V_i dN_i\right)$<br>$dU = TdS - PdV + \sum (U_i - TS_i + PV_i) dN_i - Td_{ip}S$<br>$= TdS - PdV + \sum \mu_i dN_i - Td_{ip}S$ | ref. [9] |

- $dU$: Internal energy change of the system
- $dQ$: Exchange of heat between the system and its surroundings
- $dW$: Exchange of work between the system and its surroundings
- $U_i = \left(\frac{\partial U}{\partial N_i}\right)_{dQ=0, dW=0, N_{j \neq i}}$: Partial internal energy of component $i$
- $dN_i$: Exchange of moles of component $i$ between the system and its surroundings
- $d_{ip}S$: Entropy production due to internal processes (*ip*) inside the system
- $S_i^n$ and $S_j^w$: Entropies of incoming component (nutrient) $i$ and its amount $dN_i^n$ and outgoing component (waste) $j$ and its amount $dN_j^w$ of an internal process, respectively
- $d_{ip}Q$ and $d_{ip}S^{conf}$: Heat generation and configurational entropy change due to an internal process, respectively
- $dS$: Total entropy change of the system
- $S_i = \left(\frac{\partial S}{\partial N_i}\right)_{dQ=0, d_{ip}S=0, N_{j \neq i}}$: Partial entropy of component $i$

- $V_i = \left(\frac{\partial V}{\partial N_i}\right)_{dW=0, N_{j \neq i}}$: Partial volume of component $i$
- $\mu_i = U_i - TS_i + PV_i = \left(\frac{\partial U}{\partial N_i}\right)_{S, V, N_{j \neq i}, d_{ip}S=0}$: Chemical potential of component $i$

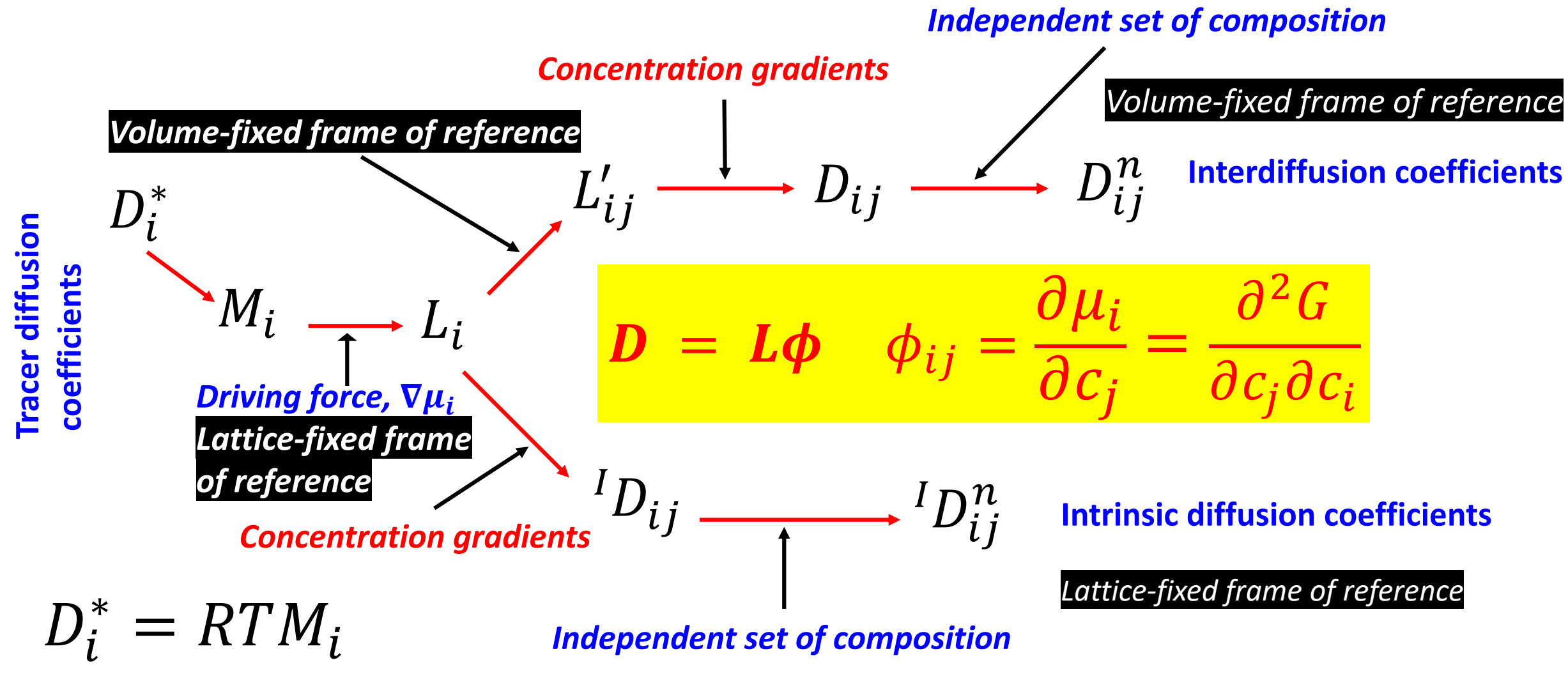

*Figure 1: Relationships among tracer diffusivity ($D_i^*$), atomic mobility ($M_i$), kinetic coefficient ($L_i$), and intrinsic diffusivities (${}^ID_{ij}$ and ${}^ID_{ij}^n$) in the lattice-fixed frame of reference; and kinetic coefficients ($L'_{ij}$) and chemical diffusivities ( $D_{ij}$ and $D_{ij}^n$) in the volume-fixed frame of reference. The relation between $M_i$ and $L_i$ is related to the diffusion mechanisms, the relation between $L_i$ and $L'_{ij}$ is related to the partial molar volume, and the relations between $L_i$ and ${}^ID_{ij}$ and between $L'_{ij}$ and $D_{ij}$ are the thermodynamic factors, i.e., the second derivative of free energy to concentration* [19,21–23].

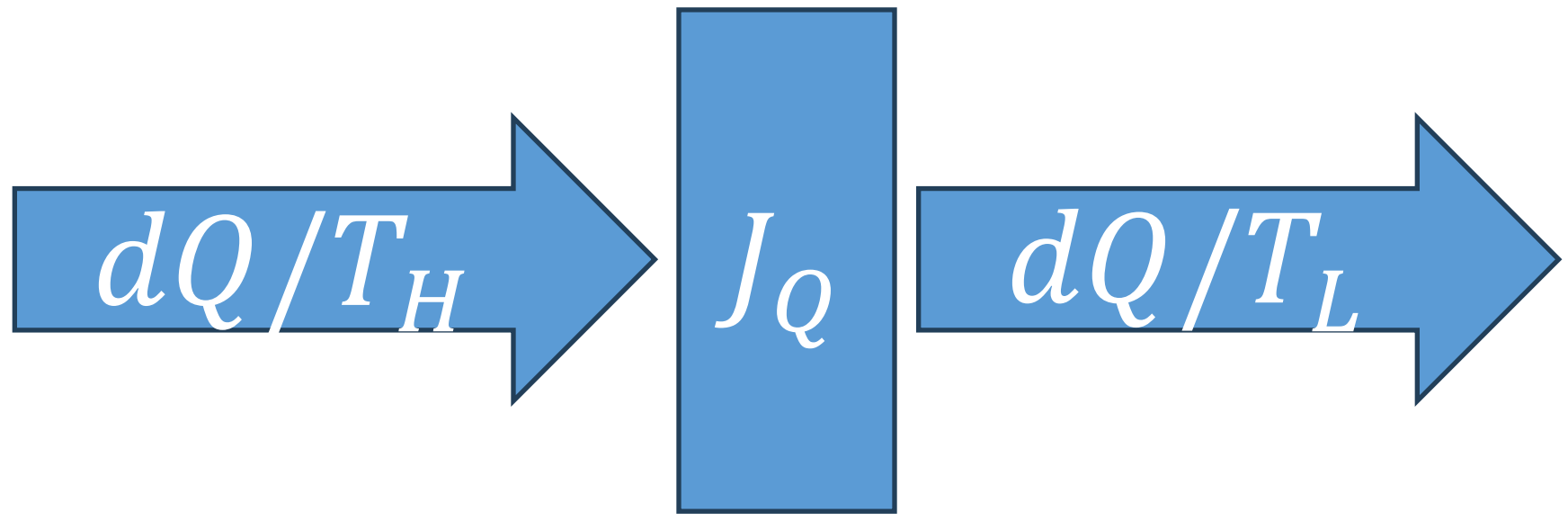

*Figure 2: Schematic heat flow from left to right.*

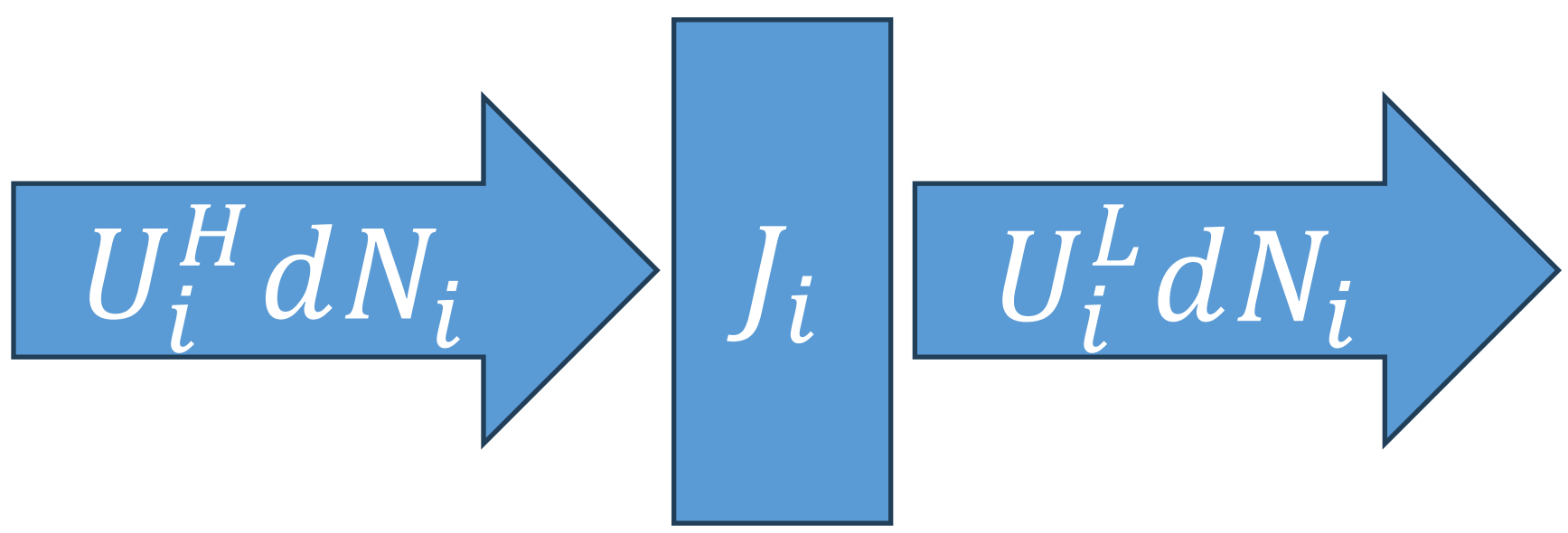

*Figure 3: Schematic mass transport of component i flow from left to right.*